\documentclass[english,a4paper]{article}
\usepackage[latin1]{inputenc}
\usepackage{epic,eepic,epsf}    
\usepackage[dvips]{graphicx}
\usepackage{epsfig}             
\usepackage{rotating}           

\usepackage{internat_immunol_cite}
  \renewcommand\citeleft{(}
  \renewcommand\citeright{)}

\begin{document}

\begin{titlepage}
%
\vspace*{2cm}
\begin{center}
{\LARGE \bf
A Possible Role of Chemotaxis in Germinal Center Formation\\[1cm]}
Tilo Beyer, Michael Meyer-Hermann, and Gerhard Soff\\
\vspace{1cm}
\end{center}
\begin{quote}
{
	Technische Universit\"at Dresden
	\\Institut f\"ur Theoretische Physik
	\\D-01062 Dresden, Germany
	\\Tel.: +49 351 463-35539
	\\Fax.: +49 351 463-37299
	\\E-mail: tilbey@theory.phy.tu-dresden.de
}
\end{quote}

\begin{abstract}
During the germinal center reaction a characteristic morphology is developed. In the framework
of a recently developed space-time-model for the germinal center a mechanism for the formation
of dark and light zones has been proposed. The mechanism is based on a diffusing differentiation signal
which is secerned by follicular dendritic cells \cite{mehe:2002}. Here, we investigate a possible influence of
recently found chemokines for the germinal  center formation in the framework
of a single-cell-based stochastic and discrete three-dimensional  model. We will also consider
alternative possible chemotactic pathways that may play a role for the development of both zones.
Our results suggest that the centrocyte motility resulting from a follicular dendritic cell-derived
chemokine has to exceed a lower limit to allow the separation of centroblasts and centrocytes.
In contrast to light microscopy the dark zone is ring shaped. This suggests that FDC-derived
chemoattractants alone cannot explain the typical germinal center morphology.
\end{abstract}

\end{titlepage}

\section{Introduction}

Germinal centers (GC) play an important role during secondary immune response
(for an overview see \cite{MacLennan:1994}). The reaction starts with the activation of B cells by antigen (Ag)
during a T cell-dependent or independent immune response \cite{Liu:1991}.
Within secondary lymphoid organs such as the spleen, tonsils, or lymph nodes the major part differentiates
into plasma cells while a small number of activated B cells---the seeder cells---migrate into the primary
follicle which now develops to a secondary follicle. There, the seeder cells develop to centroblasts \cite{Li:2000, Liu:1991}
which start to proliferate rapidly and
replace the naive B cells inside a dense network of follicular dendritic cells (FDC)
within 3--4 days \cite{Liu:1991, Camacho:1998}. At about the same time a yet unknown signal provides the
centroblasts to start somatic hypermutation \cite{McHeyzer-Williams:1993, MacLennan:1994} leading to a
diversification of antibodies (Ab) amongst the
monoclonally expanded cells \cite{Liu:1991}. Several experiments suggest that the expression of Ab is suppressed on the centroblast surface \cite{Han:1997}.
The centroblasts differentiate into smaller apoptotic centrocytes which express Ab to a higher degree \cite{Liu:1989, Liu:1994}.
To rescue centrocytes from programmed cell death they first have to interact with Ag presenting
FDCs \cite{Tew:1997, Hur:2000, Eijk:2001}. In order to do so the centrocyte has to present an Ab with high affinity for the Ag.
If the interaction is successful centrocytes are positively selected. Furthermore they have to pass a
second checkpoint where T cells test for autoimmune reaction and provide signals for further
differentiation either into plasma cells or memory cells \cite{Lindhout:1997, Eijk:2001}.
The process of somatic hypermutation together with those two
stages of selection lead to affinity maturation, i.e.~to an optimization of Abs \cite{Banchereau:1998}.

During the GC reaction a characteristic morphology occurs. When the processes of somatic hypermutation
and differentiation of centroblasts to centrocytes have started basically two specific areas
can be seen in the GC around day 5: a centroblast rich dark zone and a light zone primarily filled with
centrocytes \cite{Liu:1991, Nossal:1991, MacLennan:1994, Camacho:1998}.
Until now it has not been resolved which mechanisms lead to the separation of both cell types.
In the framework of a recent model for the GC morphology it has been shown
that a slowly diffusing signal which is secerned by FDCs and consumed by centroblasts,
indeed, leads to the intermediate development of dark and light zones
as observed in experiment \cite{mehe:2002}.

The importance of an intermediatly appearing dark zone for affinity maturation has been emphasized.
The two zones remain for at least a few days before the light zone enriches with centroblasts and the GC is
homogeneously filled with centroblasts and centrocytes \cite{Camacho:1998}.
The total cell number than decreases until the whole
GC reaction stops about 3 weeks after initiation and only a few centroblasts remain \cite{Liu:1991, MacLennan:1994}.

Today the model mentioned above \cite{mehe:2002} is the only one considering spatial aspects of the GC reaction.
Many other models deal exclusively with kinetic aspects or affinity maturation
\cite{Oprea:1997, Kesmir:1999, Nimwegen:2000, Kleinstein:2001, mehe:2001}.
In this article we will present the first three-dimensional model of the GC reaction.
The model is lattice-based and includes chemotaxis as well as a more detailed spatial resolution. 
We discuss several known 
chemotactic pathways which may play an essential role for GC formation
\cite{Bouzahzah:1996, Komai-Koma:1997, Bleul:1998, Gunn:1998, Legler:1998, Tarlinton:1998, Ansel:2000, Dubois:2001}.

\section{The Model}

To represent a GC in a spatial model we have taken into account three important concepts which
are mirrored in the following three subsections: antibodies and antigen,
the dynamics of cell-cell interactions, and the spatial distribution of cells.
In all cases we want to stay as close as possible to experimental results. Of course there
are still unknown parameters. One part of them are physiological parameters which are not yet measured or
which are not determined consistently in experiments. A second part consists of model parameters that
have no direct physiological correspondence but have to be motivated by physiological properties.
All parameter values are summarized in Table \ref{table}.
The model incorporates individual cells on a three-dimensional lattice. Each cell has its unique
set of properties described in the following sections.

\subsection{Representation of antibody and antigen}

The representation of the Ab phenotype of each cell follows the model of Perelson and Oster\cite{Perelson:1979}. In this
model a shape space, i.e.~a $N$-dimensional finite size lattice is defined in which each point corresponds
to an Ab phenotype $\Phi$. The Ag is represented by the complementary Ab position $\Phi_0$ in the shape space.
Since non-optimal Abs may also bind an Ag, the interaction between both is described by an affinity
distribution function
        \begin{equation}
        \label{eq-affinity}
        a\left(\Phi,\Phi_0\right)=\exp\left(-\frac{{\|\Phi-\Phi_0\|_1}^\eta}{\Gamma^\eta}\right)
        \end{equation}
where $\|.\|_1$ denotes the 1-norm (sum over the absolute values of the coordinates) in the shape space,
$\Gamma\approx 2.8$ is the affinity width and $\eta=2$ is the exponential weight \cite{mehe:2001}.
The affinity function is interpreted as probability of an Ab $\Phi$ to bind the Ag $\Phi_0$ ranging from $1$ for a perfect
matching Ab to $0$ for a non-binding Ab.
In this scheme hypermutation is described as a jump from an Ab position $\Phi$ to one of $2N$ possible next neighbor
Ab positions $\Phi+\Delta\Phi$. Here we do not consider key mutations which would require a description closer
to the details of somatic hypermutation. This may be provided by a genetic model including hot spots
and their implication for Abs. For a discussion of the influence of key mutations see \cite{Radmacher:1998, mehe:2001, England:1999}.
Further we do not consider effects as Ag consumption or decay \cite{Oprea:1997, Kesmir:1999}.

\subsection{Germinal center dynamics}

It exists a great variety of different pictures for the dynamics and morphological properties of GC reactions.
In our model we want to consider a rather general and simplified perspective on a representative GC. We will
concentrate on B cells and FDCs and do not model T cells, monocytes or tingible body macrophages which may
also influence the morphology.
It is known that GCs can develop their typical structure in the absence of T cells \cite{Garside:1998, Lentz:2001, Vinuesa:2000}.
The GC reaction is initiated by activated B cells which migrate into the FDC network. It has been shown
that the seeder cells have at least a low affinity for the Ag which is presented by the FDCs \cite{Agarwal:1998}.
In the model this is taken into account by using an Ab position $\Phi$ about
5 to 10 mutation steps away from the Ag position $\Phi_0$ in the shape space. This is in agreement with experimental observations
\cite{Kueppers:1993, Han:1995}.

\subsubsection{Proliferation phase}
When the seeder cells have entered the FDC network they receive a signal that induces  a fast proliferation phase
with an average cell division time of 6--7 hr \cite{Hanna:1964, Zhang:1988, Liu:1991, Liu:1994, MacLennan:1994}.
During the first three days of GC reaction about 3 seeder cells \cite{Kroese:1987, Liu:1991} increase their number to about
$10^4$ cells replacing the naive B cells inside the FDC network. The naive B cells form the mantle zone now \cite{Liu:1991}.

Since our model is single-cell based we will not simulate the proliferation in terms of a linear differential equation with constant cell
division rate. Instead, let us consider an ensemble
of equal cells in the same stage of cell cycle
and measure the time they need to complete the process of mitosis.
Qualitatively this leads to a peak centered at the mean
cell cycle time $T_P=6\,{\rm hr}$ with width $\sigma=0.6\,{\rm hr}$.
As a simple ansatz we use a Gauss-function whcih leads to a proliferation rate
        \begin{equation}
        \label{eq-Gauss}
        p(t_i)=\frac{1}{\sqrt{2\pi}\sigma}\exp\left(-\frac{{\left(t_i-T_P\right)}^2}{2{\sigma}^2}\right)
        \end{equation}
where $t_i$ denotes the \emph{eigentime} of the cell $i$. The eigentime describes an internal clock of the cell
indicating its cell cycle status and is an individual property of each cell.
We will make use of this concept throughout the model
to describe time-dependent processes such as proliferation (Sec. \ref{sec-prolif}), differentiation (Sec. \ref{sec-diff}),
or centrocyte-FDC interaction (Sec. \ref{sec-interac}).

\subsubsection{Differentiation to Centrocytes}
Some experiments show that centroblasts express low levels of Abs on their surface \cite{Han:1997}. Thus
selection is very unlikely to take place in terms of Ab affinity.
After three days of monoclonal expansion centroblasts differentiate into centrocytes which express higher
levels of Ab. Now B cells can be selected according to their affinity for the Ag presented by the FDCs \cite{Lindhout:1997}.
The first step of B cell selection within GCs is a close cell-cell interaction
between centrocytes and FDCs which takes about 1--4 hours \cite{Lindhout:1995, Eijk:1999}.

It has been shown that about at the same time when the differentiation to centrocytes starts the process of
somatic hypermutation is initiated in centroblasts \cite{McHeyzer-Williams:1993, MacLennan:1994}.

The capability of positively selected centrocytes to recycle back to centroblasts has been frequently discussed
\cite{Kepler:1993, Liu:1997, Oprea:2000, mehe:2001}. 
This is incorporated to the model using
the recycling probability $r=0.8$ \cite{Oprea:1997, mehe:2001}.
This implies that $20\%$ of all positively selected centrocytes become either plasma or memory cells.
The model does not distinguish between plasma and memory cells which are denoted as \emph{output} cells in the following.
In fact not all centrocytes that survive the first selection step become output cells due to the last selection step taking place
in interaction with T cells. So only a certain fraction $s=0.9$ of these cells will survive \cite{MacLennan:1994, Lindhout:1997, mehe:2001}.

We also include a time delay $\Delta\tau=48\,{\rm hr}$
between the onset of centroblast differentiation and the start of output cell production. All
positively selected centrocytes are recycled during this optimization phase \cite{mehe:2001, Jacob:1993, Pascual:1994},
which lasts until the onset of output cell production.

\subsection{Spatial modeling of the germinal center}\label{sec-spatial}
At first we consider the spatial resolution of the model.
We do not need to consider the details of the cell shape since we are not
primarily interested in surface properties. But we cannot neglect the
volume of the cells because centroblast are more than 10-times
larger than centrocytes (centroblast diameter $d_{\rm CB}=15\,{\rm\mu m}$ \cite{Hostager:2000}, centrocyte diameter $d_{\rm CC}=6.5\,{\rm \mu m}$
\cite{Thompson:1984, Liu:1994}).
If we allow only one cell per lattice point we have essentially two possibilities
to choose the lattice constant. On one hand,
the lattice constant can be chosen to be of the order of the centroblast diameter $d_{\rm CB}$.
This provides a well defined volume concept in the sense that one lattice point embeds a volume larger
than the volume of one cell. However, this implies that a lattice area
filled with centrocytes would mainly consist of empty space.
On the other hand, if we choose a lattice constant equal to the centrocyte diameter $d_{\rm CC}$ we do not have
problems with empty spaces but one lattice point provides not enough space for a centroblast.
As a consequence, we are led to a subcellular description that would contain more information than necessary
for the present purposes.

To solve this conflict we proceed as follows. The lattice constant $a$ is defined as
the average of the diameters of centroblasts and centrocytes. In the following the
lattice constant $a$ denotes the distance between nearest neighbor lattice points. This
is a natural unit for the discretization of parameters depending on the lattice constant
(such as cell velocities). To minimize effects of lattice anisotropy we choose a
face centered cubic (fcc) lattice which has the highest symmetry amongst regular lattices.
Second, we consider polyeders at each lattice point formed by nearest neighbor points.
Within this polyeder the volume of all cells that intersect with the polyeder
are summed up. We assume spherical cells and the lattice constant to be large enough
to exclude an intersection of cells on second next neighbor points with the polyeder.
Then, only nearest neighbors have to be considered.
A cell is allowed to move to a next neighbor point if three conditions are fulfilled:
the point is not occupied by another cell, there is enough volume remaining in the corresponding
polyeder to take up the volume of the new cell, and the polyeders belonging to the next neighbors
of the considered target point can take up the intersecting volume of the new cell.

We consider two origins for cell movement: undirectioned random movement and chemotaxis.
This is incorporated into the model by separating the cell movement into an isotropic random part and
a cell type-dependent deterministic part. In each timestep a randomly chosen
cell is highlighted to perform a motion. All next neighbor points are checked if the
cell is allowed to move to this point in the sense explained above. Then the deterministic part
is calculated as a velocity vector $\mathbf{v}$ based on chemokine concentration
gradients to \emph{free} next neighbor points (the calculation of the concentrations is
described in section \ref{sec-chemotaxis}). If the cell responds to multiple chemokines
the corresponding velocity vectors are summed up to give the resulting velocity vector $\mathbf{v}$.
This vector is projected onto the lattice via unit vectors $\mathbf{e}_n$, where $n$ denotes
all 12 possible directions of movement. The result is multiplied with the factor $\Delta t/a$
in order to convert it into a probability for moving to a lattice point via channel $n$ in
the time interval $\Delta t$:
        \begin{equation}
        \label{eq-vector}
        w_{n,{\rm deterministic}}=\mathbf{v}\cdot \mathbf{e}_n \frac{\Delta t}{a}
        \end{equation}
In an analogous way we calculate a probability for undirected random movement.
        \begin{equation}
                w_{\rm random}=v_{\rm random}\frac{\Delta t}{a N_{\rm nn}}
        \end{equation}
where $v_{\rm random}$ is the cell velocity for undirectioned movement
and $N_{\rm nn}=12$ is the number of next neighbors in the
fcc lattice.
The total probability for the movement of a cell to a neighbor point using the channel $n$
is then given by
        \begin{equation}
                w_n=(w_{n,{\rm deterministic}}+w_{\rm random})\delta_{n,{\rm free}}
        \end{equation}
where $\delta_{n,{\rm free}}$ is the Kronecker symbol for channel $n$ which is equal 1 if the channel
is free and 0 otherwise.

In order to interprete $w_n$ as probability we demand $w_n\geq 0$
and set $w_n=0$ otherwise.
As only movements to next neighbor points are allowed we have to choose the time step $\Delta t$
small enough to keep $\sum_n w_n\leq 1$. This condition depends on
the highest possible cell velocity.

Knowing the probabilities of cell movement for each \emph{free} channel, one of the available
channels is selected by using a pseudo random number. Note, that this procedure alters the effective
values for the random velocity and the chemotactic response, e.g.~in the case that all
next neighbor points are occupied by other cells both values are zero.

\subsubsection{Representation of B cells}
B cells are modeled as individual cells with several properties:
position, volume $V_i$, Ab phenotype $\Phi_i$, and eigentime $t_i$.
Each cell has a certain probability to proliferate,
if it is a centroblast, or to die by apoptosis, if it is a centrocyte, which grows according to eq.~(\ref{eq-Gauss})
when the eigentime $t_i$ reaches the average cell division time $T_P$ or lifetime $T_L$, respectively.
The eigentime is always set to zero when the cell has proliferated, i.e.~both new cells start with
eigentime $t_i=0$. The same holds true if the cell has differentiated.

\paragraph{Growth and differentiation of cells}
\label{sec-diff}
Since we included the cell volume in the model we have to explain how to
differentiate from large centroblasts to small centrocytes and back. If centroblasts differentiate to
centrocytes additional volume around the cell becomes available which is not a real
problem. But the other way round additional volume is suddenly required. In order to provide
a more realistic description of cell volume we include cell growth in the model. We use
an equation of the type
        \begin{equation}\label{eq-growth}
        \frac{dV_i}{dt_i}=\beta V_i^{\alpha} \quad ,
        \end{equation}
where $\alpha$ and $\beta$ determine the time course of the growth process.

When a centrocyte recycles back into a centroblast it starts to grow. To ensure the volume
restriction the cell may only grow if the polyeder associated to the occupied lattice point and the polyeders
associated to the next neighbor points can hold the additional volume $\beta V_i^{\alpha} \Delta t$.
If this is not possible the cell stops growing until the required space becomes available.
Note, that this process depends on the movement and growth of cells in the vicinity.
Therefore, it is important to calculate cell growth by choosing cells in a random sequence 
in each time step---as it is done for cell movement.

Addionally, the growth process of a recycled cell takes longer compared to a new cell emerging from mitosis
because the initial volume is substantially smaller. Proliferation requires the cell to have an above threshold volume.
If this is not required for the process of differentiation as well, recycled cells would preferentially differentiate
and proliferation would become a rare event. However, the multiplication of high affinity B cells through recycling is an
important feature of GC dynamics. Therefore, we forbid differentiation of a recycled centroblast until it reaches
the volume of a freshly proliferated cell.

The differentiation of centroblasts to centrocytes is assumed to occur with a constant rate after activation
by a quantum of signal molecules \cite{Choe:2000, Zhang:2001}. The influence of spatial inhomogeneous differentiation signals
on the GC morphology has been been studied before \cite{mehe:2002}. We want to exclude a spatial inhomogeneity of signal
molecules in order to avoid an interference with the chemotactic response of GC B cells. Therefore we introduce
a homogeneous density of quanta of signal molecules $\rho$. A centroblast consumes such quanta with a rate
$u=\rho\cdot 1/{\rm hr}$
proportional to this density $\rho$.
One quantum of signal molecules is assumed to contain enough molecules to induce the differentiation process.
The centroblast then differentiates with rate $1/T_D=1/(3\,{\rm hr})$.
The signal quanta are produced with rate $q$ from day 3 on.

In order to perform the differentiation process the cell 'shrinks' to the proper centrocyte volume
and acquires all characteristic properties as the ability to interact with FDCs and the initiation of the
process of apoptosis. The cell is assumed to be already able to respond to chemokines before the typical
centrocyte volume has been reached.

The constants $\alpha$ and $\beta$ in eq.~(\ref{eq-growth}) have to be specified yet.
The exponent $\alpha$ describes the volume dependence of the growth process and is determined
to be $\alpha=3/4$ according to observations made for many different organisms \cite{Kleiber:1932}.
The coefficient $\beta$ is chosen for the process of shrinking and growing separately.
In  the case of cell shrinking $\beta_s$ is determined by the time $T_S$ a centroblast needs to fully differentiate into a centrocyte.
Unfortunately this time is not known but seems to be substantially smaller than 7 hr \cite{Liu:1991,
Camacho:1998}. Therefore, we assume $T_S=3\,{\rm hr}$. In the case of cell growth we chose $\beta_g$ in respect
to the proliferation of centroblasts. The centroblast volume is doubled during the G$_1$-phase of the
cell cycle. We assume the duration of the G$_1$-phase to be of the order of $T_{G_1}=5.4\,{\rm hr}$.
It results
        \begin{equation}
        \beta_g=\frac{V_0^{1-\alpha}\left(1-2^{\alpha-1}\right)}{\left(1-\alpha\right)T_{G_1}}\approx 0.8\,\frac{{\rm\mu m}^{3/4}}{\rm hr}
        \end{equation}
for the process of growing ($V_0=(\pi /6) d^3_{\rm CB}$ denotes the maximum centroblast volume), and
        \begin{equation}
        \beta_s=-\frac{V_0^{1-\alpha}\left(1-12^{\alpha-1}\right)}{\left(1-\alpha\right)T_S}\approx-4 \,\frac{{\rm\mu m}^{3/4}}{\rm hr}
        \end{equation}
in the case of shrinking. The centroblast diameter $d_{\rm CB}$ is a little more than two times larger than the
centrocyte diameter $d_{\rm CC}$ resulting in a volume ratio of about $12:1$.

\paragraph{Proliferation of centroblasts}
\label{sec-prolif}
The proliferation of the centroblasts is described as the replacement of a large
mother cell by two daughter cells with half volume thus respecting the volume conservation
but violating the condition that only one cell can occupy one lattice point. The latter condition
is restored automatically when a lattice point in the close proximity becomes available for
one of the cells to move to.
To each new cell the eigentime zero is attributed and it restarts to grow. If the growth process
stops for a short period of time because there is not enough space available the eigentime is frozen as well.
This mirrors the influence of the growth process---the G$_1$-phase of the cell cycle---on the proliferation time.
The other phases of cell cycle are more or less of constant duration and do not depend on space restrictions
since the volume does not change.

Both daughter cells have to maintain a certain distance in order to have sufficient space
to grow. In the case of very low mobility the B cells may not reach this minimum distance
at every moment and therefore proliferation is inhibited, i.e.~the expected $10^4$ cells after three days
of clonal expansion will not be seen. This implies a lower bound for the mobility of the
B cells.

We do not consider apoptosis of centroblasts.
Experiments show that centroblasts can enter apoptosis when left without proper
surviving signals \cite{Choe:2000,Hennino:2001,Kim:1995,Lebecque:1997}. Thus we
neglect two possible effects: A time dependence and a spatial dependence of an
effective proliferation rate. Also we do not include proliferation signals \cite{Zhang:2001}.
Thus the proliferation probability $p(t)$ depends neither on the position of the cell
nor on the time course of the GC reaction. Only volume restrictions modify the proliferation rate.
The results will be discussed in this context.

\paragraph{Hypermutation}
For the selection process each B cell has its own Ab phenotype $\Phi_i$. Before the onset of
somatic hypermutation all centroblasts have the same Ab phenotype. Hypermutation
is started at the same time point as differentiation. A cell jumps to
a next neighbor point in the shape space with hypermutation probability $m=0.5$ \cite{Berek:1987, Nossal:1991}.
Hypermutation is allowed during cell division only, and both new cells mutate with the
same probability.

\subsubsection{FDC representation}
Follicular dendritic cells are represented as a soma with dendrites attached to it.
The soma is assumed to be immobile and to have a volume comparable to B cells.
The volume of the dendrites is neglected so that B cells and dendrites may occupy
the same lattice points. Each FDC has six dendrites with a length of four lattice
constants $a$ resulting in an overall length of $254\,{\rm\mu m}$ which is in good agreement with
morphological studies \cite{Szakal:1985}.
The total number of FDCs is $N_{\rm FDC}=104$ \cite{Tew:1982}. This provides enough sites for
centrocytes to interact with FDCs without saturation effects.

\subsubsection{Interaction between centrocytes and FDCs}
\label{sec-interac}
The interaction between centrocytes and FDCs takes place when the centrocyte is next neighbor
of a dendrite or the soma of an FDC or shares a lattice point with a dendrite.
We assume that the process of selection takes about 3 hours \cite{Lindhout:1995, Eijk:1999}.
During that time the centrocyte is in close contact with the FDC and therefore the centrocyte is assumed
not to move. Thereafter the affinity between Ag and the Ab phenotype of the centrocyte is calculated
according to eq.~\ref{eq-affinity}. The affinity is interpreted as probability of positive selection.
If the cell is positively selected it may become an output cell or a centroblast again.
During the optimization phase, i.e.~before the production of output cells is started
(approximately at day 5 \cite{mehe:2001, Jacob:1993, Pascual:1994}), all positively selected
B cells are recycled.

\subsubsection{Chemotaxis}
\label{sec-chemotaxis}
Every FDC soma generates a chemotactic field by secreting a chemokine into the GC.
We assume that the decay and the uptake of signal molecules by B cells is small enough to
neglect the feedback on the chemotactic field. Thus the concentration of the chemokine
remains in equilibrium during the whole GC reaction. The concentration of the chemokine
is proportional to one over distance for each FDC and is calculated once at the beginning of the
GC reaction. We further assume that the response of cells to a chemotactic
gradient $\nabla c$ is constant: $\mathbf{v}=\gamma_{a,b}\frac{\nabla c}{\|\nabla c\|}$ ($\gamma_{a,b}$
denoting the velocity of the cell type $b$ in response to a chemoattractant stemming from cell type $a$),
i.e.~the cell detects the gradient and actively moves into that direction. At every lattice point we calculate
all concentration differences to free next neighbor points resulting from all FDCs and use a vector
sum to compute the corresponding velocity vector $\mathbf{v}$ which enters eq.~(\ref{eq-vector}).
For our purposes $\gamma_{a,b}$ is the key parameter and is varied over several magnitudes to investigate
its influence on the GC morphology.

\subsubsection{Boundary and initial conditions}
We restrict the whole GC reaction to a sphere with diameter $d_{\rm GC}=345\,{\rm \mu m}$ \cite{Breitfeld:2000}
of a fully developed GC. This ensures sufficient space for slightly more than 12000 centroblast on the lattice.
No cells except output cells can leave this volume. We do not consider adhesion
or pressure of surrounding naive B cells. This assumption, or more generally, the boundary conditions
will have to be further discussed.

The starting point of our model are three activated B cells
which already entered the network of FDCs which are homogeneously distributed on one half
of the sphere mentioned above. Some experiments indicate that the FDC network is not homogeneous.
There may exist regions where FDCs present Ag and other regions where FDCs present less or almost no Ag.
In the first region centrocytes are primarily found while in the second region centroblasts are dominating \cite{Kosco-Vilbois:1997}.
In our model only the Ag containing region is take into account.

In all of our simulations we will use the same configuration of the initial Ab phenotype of the seeder cells
and the Ag held by the FDCs for reasons of comparibility.
The initial affinity of the Ab is $a(\Phi_{\rm seeder~cell},\Phi_0)=0.04$.

\section{Results}

\paragraph{GC population kinetics}
The inclusion of a differentiation signal allows various scenarios for
the GC population kinetics. Therefore at first,
we have to determine realistic time courses of the GC reaction.

The GC kinetics is strongly dependent on the production rate $q$ of the signal molecule quanta and the
boundary conditions. High production rates $q$ result in an
exponentially declining cell population. Too small production rates $q$ imply
a domination of proliferation over differentiation and the cells fill the whole
lattice until the proliferation is inhibited by space limitations.
This leads to a durable dynamic equilibrium with a stable B cell population
(data not shown).
For production rates between these two extreme scenarios
the time course of the B cell population becomes realistic.
At the beginning B cells fill the whole lattice.
After a certain time (depending on the signal production rate $q$) the
signal molecule density increases and more centroblasts start to
differentiate.
This results in a controled reduction of the
centroblasts population (Fig.~\ref{fig-reaction}).

The typical GC reaction in the model shows a nearly exponential increase of the centroblast population during
the first $72\,{\rm hr}$ (Fig.~\ref{fig-reaction}). Then the differentiation process starts and the centrocyte population
increases. The peak of the reaction is reached at about $96\,{\rm hr}$ in good agreement with experiment \cite{Vinuesa:2000}.
The total population then stabilizes on
a plateau for some days before it declines rapidly
entering a long lasting low level reaction
with a very slowly decreasing B cell population.

The time course of such a GC reaction is only slightly dependent on the differentiation rate $1/T_D$,
describing the rate with which activated centroblasts differentiate into centrocytes. The differentiation
rate has basically to be large enough to guarantee a declining GC population in the late stages of the reaction \cite{mehe:2001}.
Note, that observable
differentiation rates are effective rates which already include the effect of signal molecules.

\paragraph{FDC-derived chemoattractant for centrocytes}
For our purpose we define the appearance of a light zone as a cluster of cells in which centrocytes dominate.
The dark zone is analogously defined.
To achieve a separation of centrocytes and centroblasts in a GC the FDCs may selectively attract
centrocytes but not centroblasts \cite{Komai-Koma:1997}. To test this hypothesis we simulate the GC reaction
with a chemotactic response of centrocytes only.

\begin{figure}[p]
 \input{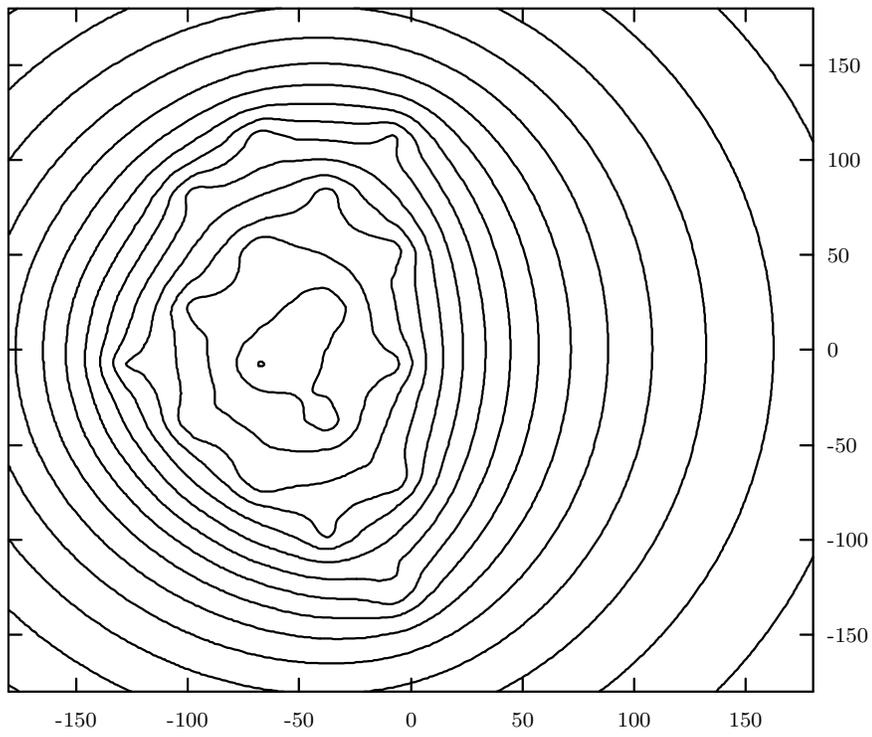}
 \caption{\label{fig-taxisfield}The equipotential lines of the
        chemotactic signal concentration.
        From outside to inside every line represents
        an increase of $0.5$ of the chemokine concentration in
        arbitrary units.}
\end{figure}

A light zone is observed if the centrocyte velocity resulting from the chemokine allows
the cells to leave the centroblast population before they die by
apoptosis.
A dark zone develops if this velocity is
high enough to allow centrocytes to leave the centroblast population before new
centrocytes arise from centroblast differentiation,
thus generating a centroblast dominated area.
For very small chemotactic coefficients $\gamma_{\rm FDC,CC}$ no zones occur at all (data not shown).
Large chemotactic coefficients $\gamma_{\rm FDC,CC}$ result in the formation of a light and dark zone. While the light zone
consists of a dense cluster of centrocytes with only few centroblasts, the dark zone is formed in an
asymmetric ring like structure of less dense packed centroblasts (Fig.~\ref{fig-fdc}). This can be understood
by assuming a quasi-stationary situation in which cell numbers do not
significantly change. Then the
centrocytes get entrapped in the local minima of the chemotactic field reproducing the equipotential lines of the
signal molecule concentration (Fig.~\ref{fig-taxisfield}).
Note, that not all cells can achieve local
minima since volume restrictions have to be respected. The unphysiological ring structure of the dark zone
is caused by the fact that the centroblasts have no attractor and behave like a gas inside the GC sphere.
With intermediate chemotactic coefficients the zones appear less clearly and later (Fig.~\ref{fig-btaxis} left column).
The resulting deterministic velocity critical for the formation of light zones is of the order of several
$\mu m$ per $min$.

\paragraph{Required motility of centroblasts}
The reaction also requires that the undirected
movement of the centroblasts is fast enough (with
a random velocity $v_{\rm random}\approx 2\,{\rm \mu m/min}$) in order
to allow the centrocytes to escape the centroblast dominated areas.
For random velocities which are significantly below this threshold no light zone
develops as long as the centroblast population forms a dense packed cluster.
Only when the density of centroblasts
declines a light zone occurs for a short period of time (data not shown).
Remarkably, the necessary random velocity shows no significant dependence on the chemotactic coefficient
$\gamma_{\rm FDC,CC}$.
The undirected movement of centroblasts is necessary to
offer the possibility to centrocytes to use small gaps for movement.
Higher centrocyte velocities have basically no effect if the space for movement is lacking.
On the other hand a faster undirected movement of centroblasts
may offer enough dynamically produced gaps to ensure the mobility of centrocytes. However, the directed centrocyte
movement may be to slow to use the available space.
We conclude that a reasonable relation of centroblasts and centrocytes mobility is necessary
to allow the separation of both cell types.

In the cases where dark and light zones appear it can be observed that the centrocytes can leave the centroblast
dominated areas.
This results from the high mobility and small diameter of centrocytes which use every available space around
the large centroblasts to perform a movement in direction of the FDC network.
However, they do not
replace all of the centroblasts inside the FDC network.
This is in parts related to the process of recycling which acts as an additional source
of centroblasts in the FDC network.
In addition, centroblasts rarely leave the FDC network due to their
relatively weak mobility.
In conclusion, dark zones with very low numbers of centrocytes
and light zones with relatively high fraction of centroblasts are generated.

\paragraph{Chemoattractant for centroblasts}
In a second step we let the centroblasts also respond to the FDC chemokine with equal, higher and weaker response
compared to the centrocytes. In all three cases there exists neither a dark nor a light zone during the whole GC reaction
(Fig.~\ref{fig-btaxis} right column) except if the chemotactic response of the centroblasts is weak enough
so that the undirected movement dominates.
Then, the centroblasts behave similar to a free gas as in the scenario
where only centrocytes respond to the chemokine.
In the case of weak
chemotactic response of centroblasts ($\gamma_{\rm FDC,CB}$) and strong chemotactic
response of centrocytes ($\gamma_{\rm FDC,CC}$) a very small symmetric ring of centroblasts occurs
(Fig.~\ref{fig-btaxis} right column, day 7).

\paragraph{Mantle zone-derived chemotactic signals}
In order to avoid the ring structure
we investigate the mantle zone as an alternative source for a chemokine that may act on centrocytes. We include
a preformed mantle zone of a GC which is polarized, i.e.~the mantle zone is thicker on the side
of the FDC network. Similar to the FDCs we let the cells in the mantle zone segregate a chemokine. As before
this is reflected in a stationary configuration. This signal alone is sufficient to form light and dark zones
(Fig.~\ref{fig-mz}). The dark zone is even more physiological than with FDC-derived chemokine but begins
to penetrate the FDC network from day 8 on while the light zone moves towards the boundary of the FDC network
around day 9 (Fig.~\ref{fig-mz}). This is due to the new location of the local minima of the chemotactic field
situated at the outer boundary of the FDC network. Centrocytes are densely packed in these local minima. When their
number begins to decline they do not longer extend to the FDC network. The centroblasts behave like a gas
and are dispersed over the free space including the part of the FDC network opposite to the mantle zone.
This affects also the late stage of the GC reaction. The B cell population declines much faster than with
FDC-derived chemokine and therefore generates less output cells (data not shown).

\paragraph{Combined chemotactic signals}
When the centrocytes respond to signals from FDC \emph{and} mantle zone cells, a light zone and a dark zone
can be observed (Fig.~\ref{fig-mz-fdc}). The light zone stays within the FDC network during the whole reaction.
The dark zone neither is ring shaped nor
penetrates the FDC network. The GC morphology is closer
to observed morphologies than with FDC-derived chemokine alone. Also the GC kinetics stays in good agreement
with experiment (data not shown). Only the differentiation rate $1/T_D$ has to be adjusted to restore the late stage
of the reaction. Otherwise,
the B cell population would decline faster than with FDC-derived chemokine only, because
the FDC-centrocyte interaction is slightly reduced when some of the centrocytes are entrapped at the boundary of
the FDC network.

\begin{figure}[p]
 \input{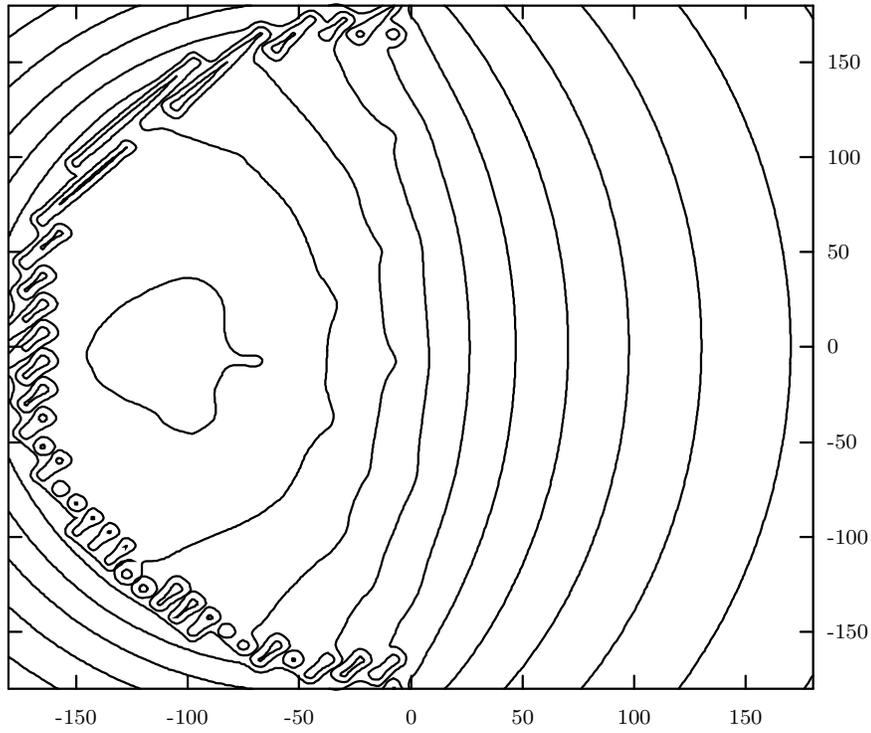}
 \caption{\label{fig-chemofield}The equipotential lines of the effective chemokine concentration.
        Isolines of the FDC-derived
        plus the mantle zone-derived chemokine concentrations are shown.
        The relative strength of the signals is $1:2$. From outside to inside every line represents
        an increase of $0.5$ of the chemokine concentration in arbitrary units (same as in Fig.~\ref{fig-taxisfield}).
        The global minimum is shifted to the mantle zone but remains within the FDC network.}
\end{figure}

The relative strength of the chemotactic response
must be tightly balanced. The velocity in response to the mantle zone-derived chemokine has to be
twice the strength compared to the response to FDC-derived chemokines. Higher values lead to structures like in Fig.~\ref{fig-mz} and
smaller values to structures like in Fig.~\ref{fig-fdc}. To explain this behavior we
plot the equipotential lines of the resulting chemotactic field (Fig.~\ref{fig-chemofield}).
The shift of the isolines towards the mantle zone compared to Fig.~\ref{fig-taxisfield} can clearly be seen.
We want to emphasize that the B cells in the mantle zone are not necessarily
the source for
the chemokine. Any other cell type with similar spatial distribution leads to the same result.

\paragraph{Affinity maturation}
If we investigate the affinity of the output cells we recognize that we have approximately $60\%$ output cells
with high affinity ($a(\Phi,\Phi_0)\geq 0.8$) and about $30\%$ medium affinity cells ($0.4\leq a(\Phi,\Phi_0)\leq 0.8$)
(Fig.~\ref{fig-affinity}). The plot shows the
fraction of corresponding numbers of output cells
integrated over time intervals of $6\,{\rm hr}$ and
reflects the quality of produced output cells. The small numbers of produced output cells during those short time intervals
lead to huge statistical fluctuations. Remarkably, the ratio between medium and high affinity cells is
reversed after two weeks. At the beginning of the output production most cells have still
low affinity for the Ag. Then, the fraction of medium and high affinity cells increases. After two weeks the
fraction of high affinity cells increases further while the fraction of medium affinity cells is declining.
This behavior is in good quantitative agreement with experiment \cite{Smith:1997}.

Interestingly, this behavior seems to be independent of the GC morphology as long as the general GC kinetics remains
unaffected. In contrast, the number of output cells changes for different morphologies. The output quantity depends on the
probability for centrocytes to interact with FDCs. Some morphologies of GC reaction induce volume constellations that
inhibit the movement of centrocytes and in this way reduce this interaction probability.
However, the fact that the quality
remains almost the same (data not shown), at first sight, seems to be in contradiction to
results of Meyer-Hermann \cite{mehe:2002}.
But we use another indicator for the quality of the GC reaction.
Considering the time course of the time integrated output quality instead, the GC morphology, indeed, influences
the total output quality (data not shown).
This value memorizes the total output production and is therefore sensitive to the effectivity
of affinity maturation in the early phase (around day 6) of the GC reaction which
is inhibited by the absence of zones \cite{mehe:2002}.

\paragraph{Robustness of the model and the results}
The model results are stable against small variations of most of the experimentally unknown parameters.
This applies especially to the parameters
of cell movement ($v_{\rm random}, \gamma_{a,b}$). The most sensitive parameter is the
production rate of the signal molecule quanta $q$. Its value determines the duration of the plateau phase
in the GC kinetics and variations
of more than 10\% cause exponentially declining reactions
or persistent plateau phases, respectively.
Also the total size of the GC represented by $d_{\rm GC}$
influences the GC kinetics. The plateau phase of the reaction is reached
when the proliferation is inhibited due to limited space.
This results in an increased number of
centroblasts. When the signal production rate $q$ is adjusted the GC kinetics in Fig.~\ref{fig-reaction} is
restored on a higher level of
cell numbers. In order to fit the cell numbers to
experiment \cite{Liu:1991}, the value for the
diameter of the GC is fixed to $d_{\rm GC}=345\,{\rm \mu m}$.

Despite the stochastical nature of the model, the results are reproducible without significant statistical
variations. The only exception are the details of the affinity maturation (Fig.~\ref{fig-affinity})
due to small cell numbers.

Also the morphology is independent of the details of the FDC distribution. An expansion of the FDC network
slightly beyond the boundary of the GC does not alter the results in general.

\section{Discussion}

In the present study we enlarged our previous model for the GC morphology in order to investigate
the influence of chemotaxis on the GC formation. To this end it was necessary to develop
a more detailed description of cell mobility. Besides a minimization of lattice-anisotropy effects,
this, especially, includes a self-consistent cell-volume concept on a discrete regular lattice without
going into the details on a subcellular level. We included cell growth and shrinking, as well
as cell differentiation.
The resulting model is the first three-dimensional attempt to simulate GC cell population
dynamics and GC morphology. It should, therefore, provide a more realistic description of
processes involved in the GC development.

Within our model the different cell volumes of centroblasts and centrocytes strongly influence
their mobility. We observe in the simulations that small centrocytes are able to move in the
environment of large centroblasts even when the latter ones are densely packed. The centrocytes
find small gaps and slip through the centroblasts. The mobility of the centroblasts is
inhibited by their larger volume. The restrictions on the total GC volume inhibit centroblasts growth
and in this way centroblast proliferation. All of these observations in the GC simulations
are quite realistic effects which have an equivalent in real GCs, and which seem to be important
on the way towards a realistic picture of moving cells during the GC reaction. We consider therefore our
model to be suitable for studying possible effects of chemotactic signals.

\begin{figure}[p]
 \begin{center}\includegraphics[width=12cm]{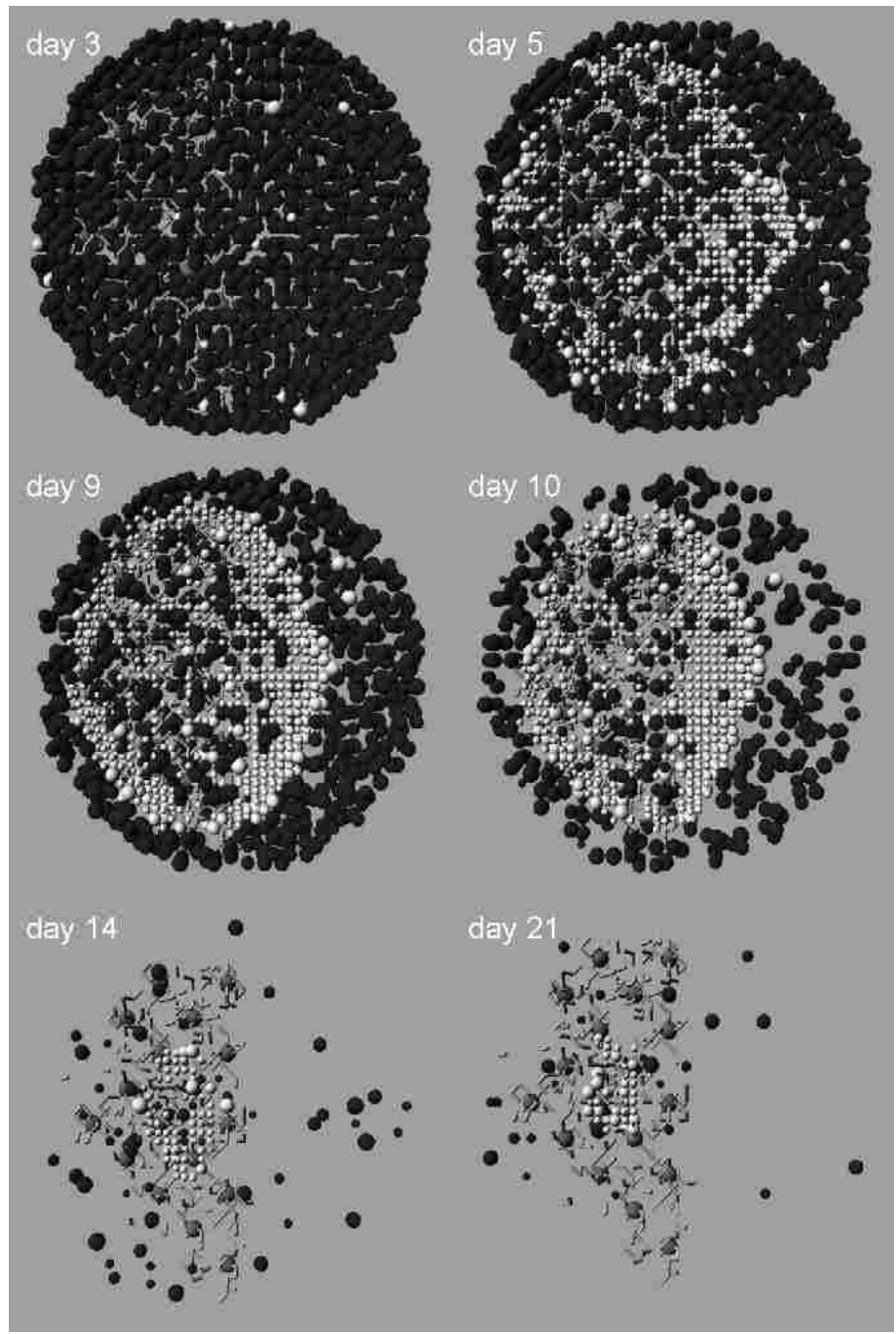}\end{center}
 \caption{\label{fig-fdc}A $2a$ thick slice of
        the germinal center (centrocytes - white, FDC - grey,
        centroblasts - black). Only centrocytes respond to
        the chemokine secerned by the FDCs with $\gamma_{\rm FDC,CC}=13\,{\rm \mu m/min}$
        (for all other parameters see Table \ref{table}). From day 5 on a light zone can be observed.
        It remains stable for at least five days but never exclusively consists of centrocytes.
        There are always centroblasts present which are mostly recycled centroblasts (not distinguished in the
        figure). The dark zone has a ring like shape. The last figure
        shows the GC after 3 weeks with only small numbers
        of centroblast and centrocytes remaining.}
\end{figure}

\begin{figure}[p]
 \begin{center}\includegraphics[width=12cm]{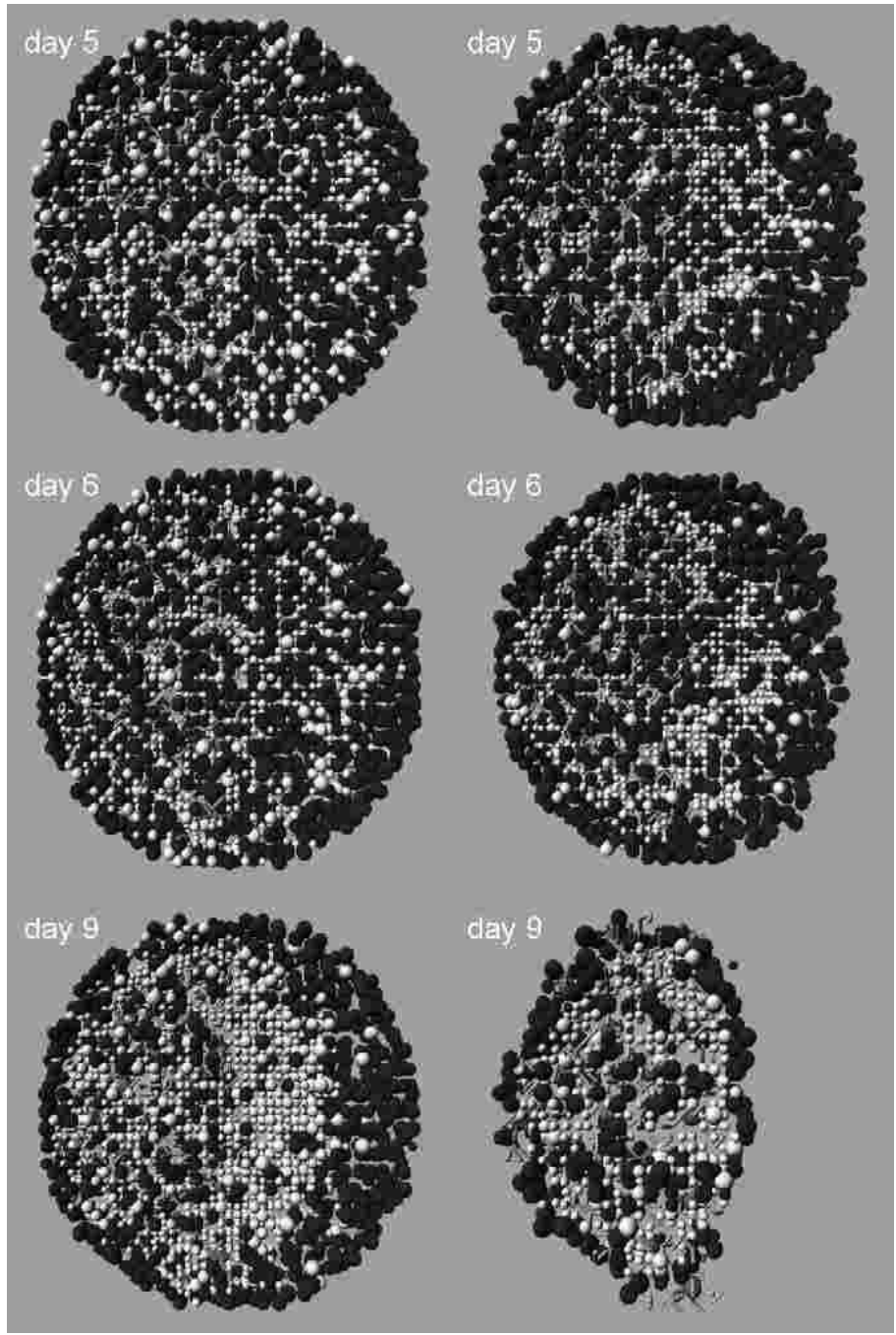}\end{center}
 \caption{\label{fig-btaxis}A $2a$ thick slice of
        the germinal center (centrocytes - white, FDC - grey,
        centroblasts - black). The left column shows a GC reaction with an intermediate chemotactic response
        of centrocytes to a FDC-derived chemokine $\gamma_{\rm FDC,CC}=1\,{\rm \mu m/min}$. The light zone is less clearly
        formed compared to Fig.~\ref{fig-fdc} day 5-10. Only, when the density of the cells is reduced
        around day 7 a more pronounced light zone can be observed. The right column shows a GC with centroblasts responding
        to a FDC-derived chemokine $\gamma_{\rm FDC,B}=2\,{\rm \mu m/min}$. Almost no separation of centrocytes and centroblasts
        can be seen. Only a small dark zone appears. All other
        parameters remain as in Table \ref{table}.}
\end{figure}

First, we investigated the influence of a FDC derived chemoattractant \cite{Bouzahzah:1996, Komai-Koma:1997}
on the GC morphology. This signal is sufficient to separate centroblasts and centrocytes but can not explain
the formation of the characteristic GC structures. If the chemotactic response of the centrocytes and the
random movement of the centroblasts exceeds given values ($\gamma_{\rm FDC,CC}=4\,{\rm \mu m/min}$ for centrocytes
and $v_{\rm random}=2\,{\rm \mu m/min}$ for centroblasts, respectively) a light and a dark zone develop.
But in contrast to experiment the centroblasts form an asymmetric ring like structure (Fig.~\ref{fig-fdc}).
For smaller values of the cell mobility (about one order of magnitude) the zones are developed later in the
GC reaction when the total density of cells becomes small enough (Fig.~\ref{fig-btaxis}). For significantly
reduced cell mobility no separation into light and dark zone occurs at all. The corresponding velocities of
both cell types are within physiological relevant values and are comparable to relatively slowly moving cells
\cite{Boll:1992, Murray:1992, Felder:1994, Masellis-Smith:1996, Niggemann:1997, Friedl:2001}.

Interestingly, chicken GCs seem to have such a ring structure suggesting that FDC-derived chemotaxis acting
on centrocytes is adequate to describe their morphology \cite{Yasuda:1998}. In contrast, in mammalian GCs
other or addional mechanisms cause the formation of dark and light zones. To test this hypothesis the mantle
zone was taken into consideration as source of a chemokine acting on centrocytes. Again a separation into
light and dark zone can be observed. The dark zone is sickle-shaped and the light zone is shifted within
the FDC network towards the source of the chemoattractant. This causes an inhibited FDC-centrocyte interaction
resulting in reduced numbers of recycled centroblasts and output cells.

In addition we investigated centroblasts responding to FDC-derived chemokines. Only a small ring shaped
dark zone can be achieved for small values of the chemotactic coefficient $\gamma_{\rm FDC,CB}=2\,{\rm \mu m/min}$ and
a more intense chemotactic response of centrocytes $\gamma_{\rm FDC,CC}=10\,{\rm \mu m/min}$ (Fig.~\ref{fig-btaxis}).
The ring of centroblasts is now symmetric in contrast to the results when the centroblasts do not respond to
the FDC-derived chemoattractant. The asymmetric ring structure is restored for significantly reduced chemotactic
response when the random motility dominates (data not shown).
It is known that the seeder B cells enter the primary follicle via chemotaxis provided by the BLC - CXCR5 pathway
\cite{Ansel:2000, Gunn:1998, Legler:1998, Shi:2001}. Our results suggest that centroblast only weakly respond
to FDC-derived chemokines during the GC reaction. In agreement with this GC B cells showed no response to BLC
\cite{Bowman:2000}, SDF-1$\alpha$ \cite{Bleul:1998, Bowman:2000}, SLC, and MIP-3$\alpha$ \cite{Bowman:2000},
and a weak response to rC5a \cite{Ottonello:1999}.

\begin{figure}[p]
 \begin{center}\includegraphics[width=12cm]{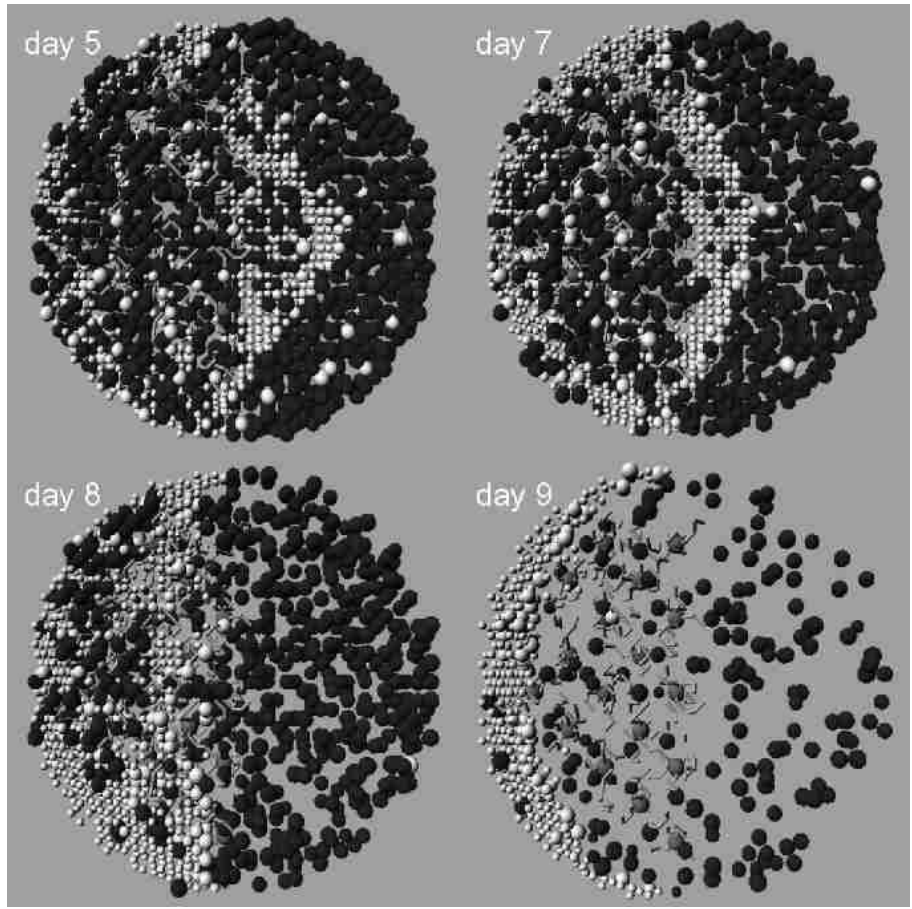}\end{center}
 \caption{\label{fig-mz}A $2a$ thick slice of
        the germinal center (centrocytes - white, FDC - grey,
        centroblasts - black). A GC reaction with mantle zone-derived chemokines only.
        Dark and light zone can be distinguished. The dark zone is sickle shaped instead of ring shaped (Fig.~\ref{fig-fdc} day 5-10)
        but begins to penetrate the FDC network from day 8 on. The light zone is shifted towards the boundary of
        the FDC network after day 9 and the GC reaction stops very soon because positive selection and recycling
        are suppressed (not shown).}
\end{figure}

We also analyzed alternative sources for the chemokines. If the mantle zone B cells or cells with
a similar distribution secrete a chemokine acting as attractant for centrocytes, dark and light zones are
formed (Fig.~\ref{fig-mz}). The separation of dark and light zone is more physiological but results in a shorter GC reaction
because the light zone shifts to the boundary of the FDC network while the dark zone enters it. This may
be related to the model assumptions.
We did not include the dynamics of the mantle zone thus neglecting a possible
movement of the minima of the chemotactic field relative to the FDC network. The duration
of the GC reaction can only be prolonged if unrecycled centroblasts are attributed a longer lifetime to.
But the quality of the output cells still remains too low (data not shown).

\begin{figure}[p]
 \begin{center}\includegraphics[width=12cm]{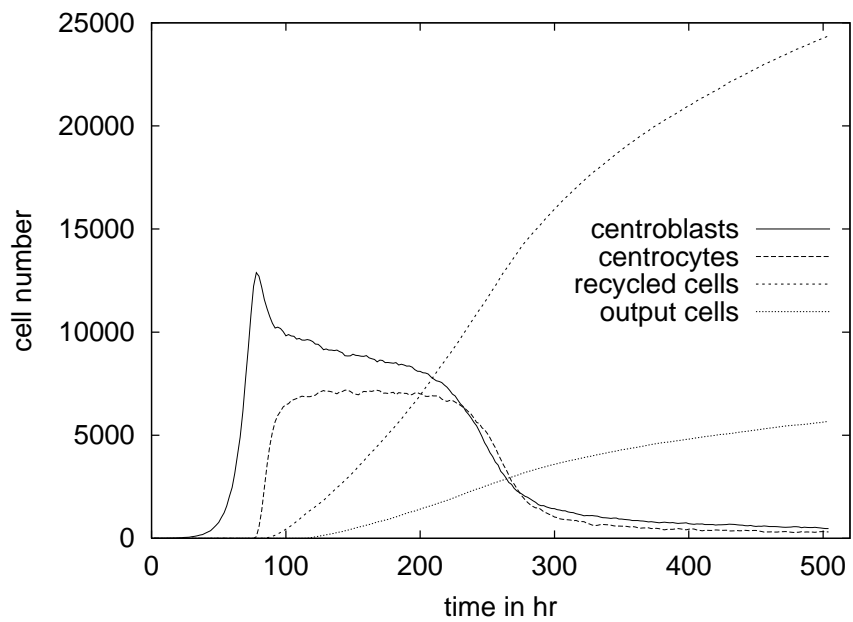}\end{center}
 \caption{\label{fig-reaction}Different cell populations during the GC reaction. Recycled and output cells
        are shown as time integrated cell numbers.  For centroblast and centrocytes
        cell numbers at given time points are plotted.
        After $72\,{\rm hr}$ the differentiation into centrocytes starts and the centrocyte population increases rapidly.
        The total B cell number (centroblast and centrocytes) has a peak around day 4 ($=96\,{\rm hr}$) followed by
        a plateau. After 10 days the populations is declining.}
\end{figure}

\begin{figure}[p]
 \begin{center}\includegraphics[width=12cm]{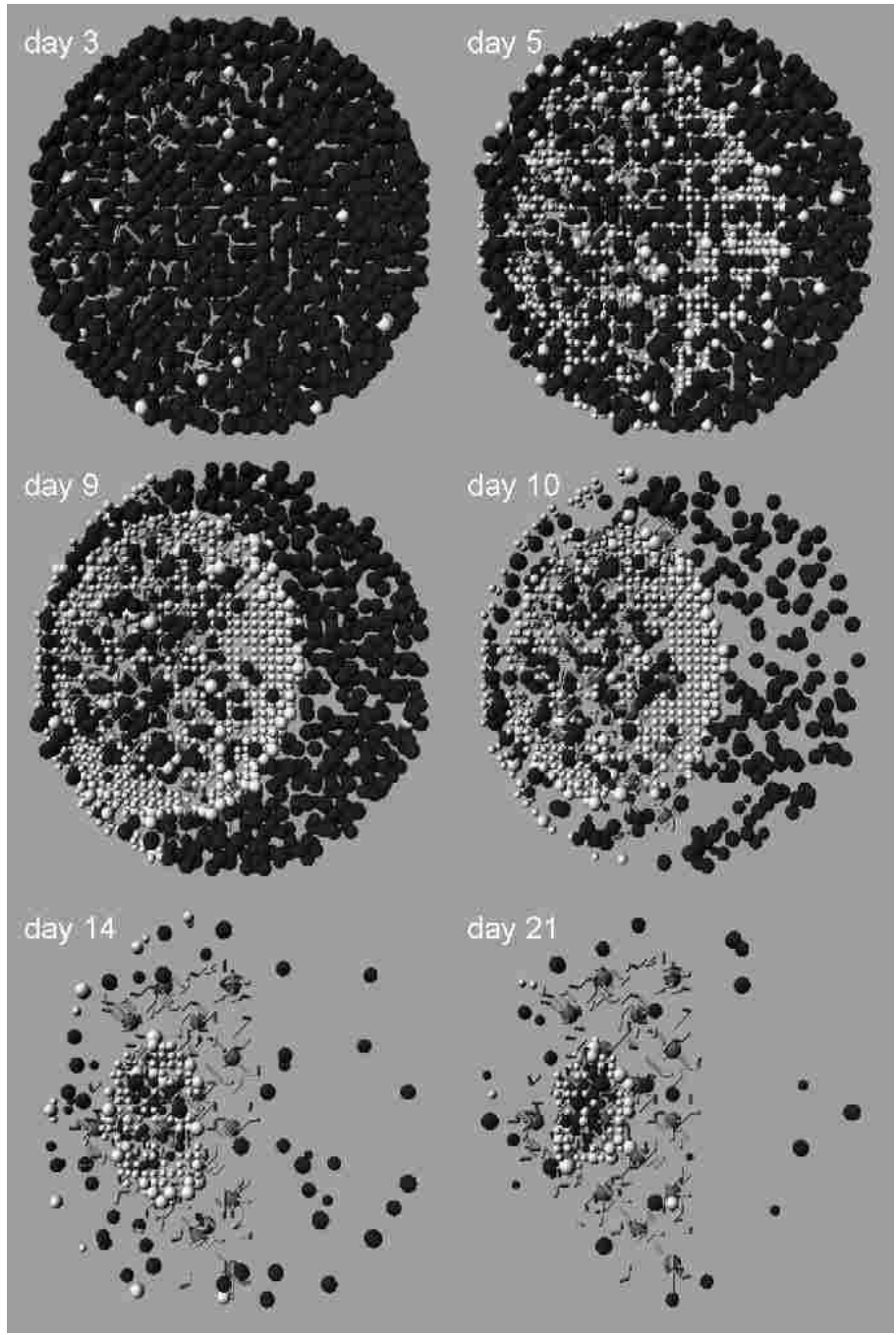}\end{center}
 \caption{\label{fig-mz-fdc}A $2a$ thick slice of
        the germinal center (centrocytes - white, FDC - grey,
        centroblasts - black). The centrocytes respond to
        chemokines from FDCs $\gamma_{\rm FDC,CC}=7\,{\rm \mu m/min}$ and mantle zone cells $\gamma_{\rm MZ,CC}=13\,{\rm \mu m/min}$.
        We can observe a sickle shaped dark zone and a light zone. The ring structure of the dark zone
        is clearly reduced (compare Fig.~\ref{fig-fdc} day 5-10). The dark zone does not penetrate the FDC network, and the
        centrocytes stay within the light zone (compare Fig.~\ref{fig-mz} day 8 and 9). The differentiation rate
	was adjusted to $1/T_D=1/3.5\,{\rm hr}$ (all other parameters see Table \ref{table}).
        }
\end{figure}

\begin{figure}[p]
 \begin{center}\includegraphics{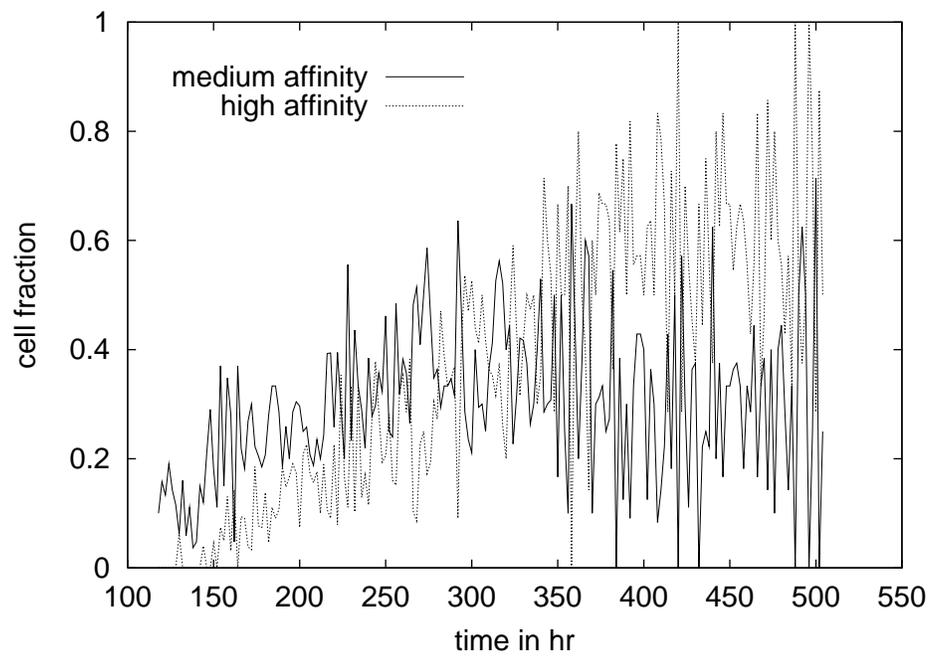}\end{center}
 \caption{\label{fig-affinity}The fraction of medium and high affinity output cells in small time windows
        of $6$ hours.
        Medium affinity is defined as $0.4\leq a(\Phi^*,\Phi_0)\leq0.8$
        and high affinity as $0.8< a(\Phi^*,\Phi_0)\leq 1.0$. $\Phi^*$ denotes all $\Phi$ in the given
        intervals and $\Phi_0$ the position of the Ag. The output production begins after 5 days (120\,{\rm hr}).
        }
\end{figure}

When centrocytes respond to both chemokines---the mantle zone-derived and FDC-derived chemokine---a longer
lasting light zone can be observed which remains inside the FDC network. The dark zone is sickle shaped and
does not expand into the FDC network~(Fig.~\ref{fig-mz-fdc}). This is the most realistic scenario we could
generate. The GC kinetics and the affinity maturation are similar to the scenario with FDC chemokine alone
(Fig.~\ref{fig-reaction} and Fig.~\ref{fig-affinity}) while the unphysiological ring structure of the dark zone
turns into a more realistic sickle shaped dark zone. This result still does not reproduce the observations
of light microscopy \cite{Liu:1991, Schaerli:2000, Steininger:2001, Verbeke:1999} suggesting that 
other mechanism are necessary to generate the typical GC morphology.
One possible mechanism is a diffusing differentiation signal secerned by the FDCs \cite{mehe:2002}.

In general we achieved physiologically realistic GC population kinetics for
the different scenarios of the GC morphology
{Fig.~\ref{fig-reaction}}. In the late stages of the reaction when
low numbers of GC B cells remain
differences occur affecting the number of output cells. This results in different total average affinities
of output cells depending on
the various GC morphologies. However, 
the quality of output cells produced at the end of the reaction are comparable
provided that the number of output cells is large enough.
The distribution between high, medium, and low
affinity cells (Fig.~\ref{fig-affinity}) quantitatively mirrors experimental results very well \cite{Smith:1997}.
We can conclude that the morphology of the GC mainly influences the quantity and not the quality of output cells.

One assumption of the model is that centrocytes respond to a chemokine with constant velocity only detecting
the direction of the chemotactic gradient. We addressed the question if a linear dependency of the velocity
on the chemokine concentration gradient ($\mathbf{v} \sim \nabla c$), would change the results. Indeed, the
results can not be restored within reasonable parameter values, i.e.~velocities. Centrocytes far away from
the FDC network have the smallest velocities and at the same time the longest way to go. However, increasing
the over all velocity in response to the chemotactic gradient means, especially, also increasing the velocity of cells close
to the FDC network reaching unphysiological high values. If the chemotactic response was linear to the
concentration itself one would get almost the same result with a slightly flatter velocity distribution (one
over distance compared to one over distance squared). Taken together, this suggests that centrocytes use the
chemokine concentration mainly
to detect the direction of movement and have a more or less constant velocity in response to the
chemoattractant.

One may think about a possible influence of cell-cell adhesion.
It is known that GC B cells and FDCs form
clusters in vitro \cite{Kosco:1992}. The cooperation of chemotaxis and differential adhesion for the
separation of cells has been studied in other systems \cite{Jiang:1998}.
The major difference to this study is that centroblasts differentiate into centrocytes and vice versa
while in the model of Jiang et al.~the cell fractions remain constant. In addition, the cells have no gaps
between them while in our model small gaps are necessary to allow the centrocytes to move towards the FDC.
It would be interesting to investigate if adhesion could substitute the boundary condition,
namely the restriction of
cell movement to a sphere, and thus circumvent the gas behavior of the centroblasts.

We conclude, that FDC-derived chemokines acting on centrocytes lead to a separation
of centroblasts and centrocytes in GC reaction. However, the
GC morphology as observed in mammalians is not correctly reproduced.
This problem persists even using a combined centrocyte response
to mantle zone-derived and FDC-derived chemokines. These results suggests
that chemotaxis for its own is not sufficient to induce a realistic development of
dark and light zones. The present model points towards an additional role either
of cell-cell adhesion or of cells surrounding the GC.

\bibliographystyle{internat_immunol}
\bibliography{literatur}

\clearpage

\begin{table}
 \begin{tabular}{|p{7.0cm}|r|p{2.5cm}|}\hline
 parameter                      &value          &source\\
 \hline \hline
 {\bfseries time resolution $\mathbf{\Delta t}$}&$\mathbf{10^{-2}\,{\rm hr}}$   &    \\
 {\bfseries lattice constant $\mathbf{a}$}      &$\mathbf{10.6\,{\rm \mu m}}$  &    \\
 centrocyte diameter     $d_{\rm CC}$           &$6.5\,{\rm\mu m}$    &\cite{Thompson:1984, Liu:1994}  \\
 centroblast diameter    $d_{\rm CB}$           &$15\,{\rm\mu m}$     &\cite{Hostager:2000}\\
 germinal center diameter        $d_{GC}$               &$345\,{\rm \mu m}$    & \cite{Breitfeld:2000}                   \\
{\bfseries random velocity $\mathbf{v_{\rm random}}$}   &$\mathbf{2\,{\rm \mu m/min}}$   &    \\
 \hline
 number of seeder cells                         &$3$            &\cite{Liu:1991}     \\
 number of FDCs         $N_{\rm FDC}$           &104            &\cite{Tew:1982}     \\
 proliferation time     $T_P$                   &$\,{\rm 6hr}$          &\cite{Hanna:1964, Zhang:1988, Liu:1991, Liu:1994, MacLennan:1994}             \\
 {\bfseries proliferation time width} $\mathbf{\sigma}$  &$\mathbf{0.6\,{\rm hr}}$        &                    \\
 differentiation time              $T_D$        &$3\,{\rm hr}$  &\cite{Liu:1991, mehe:2001}          \\
 {\bfseries production of differentiation signal molecule $\mathbf{q}$}  &$\mathbf{900/{\rm hr}}$       &       \\
 centrocyte lifetime $T_L$                      &$6\,{\rm hr}$          &\cite{Liu:1991}     \\
 duration of selection process                  &$3\,{\rm hr}$          &\cite{Lindhout:1995,Eijk:1999}\\
 growth exponent      $\alpha$                  &$0.75$         &\cite{Kleiber:1932} \\
 {\bfseries duration of growth $\mathbf{T_{G_1}}$}       &$\mathbf{5.4\,{\rm hr}}$       &                    \\
 {\bfseries growth coefficient $\mathbf{\beta_g}$}       &$\mathbf{0.8\,{\rm\mu m}^{3/4}/{\rm hr}}$   &       \\
 {\bfseries duration of differentiation process $T_S$}& $\mathbf{3\,{\rm hr}}$          &       \\
 {\bfseries shrink coefficient $\mathbf{\beta_s}$}       &$\mathbf{-4\,{\rm \mu m}^{3/4}/{\rm hr}}$    &       \\
 \hline
 recycling probability $r$                              &$0.8$          &\cite{mehe:2001}    \\
 affinity width     $\Gamma$                    &$2.8$          &\cite{mehe:2001}    \\
 exponential weight $\eta$                      &$2$                    &\cite{mehe:2001}    \\
 shape space dimension $N$                      &$4$            &\cite{Perelson:1979, mehe:2001}    \\
 hypermutation probability $m$                  &$0.5$          &\cite{Berek:1987, Nossal:1991}\\
 time delay between start of differentiation and production of output cells $\Delta\tau$        &$48\,{\rm hr}$ &\cite{mehe:2001}\\
 \hline
 \end{tabular}
 \caption{\label{table} Summary of model parameters. Model parameters in bold are non-physiological or corresponding experimental results are
        inconsistent or not available.}

\end{table}

\clearpage

\setlength{\unitlength}{1cm}

\clearpage

\clearpage

\clearpage

\clearpage

\clearpage

\clearpage

\clearpage

\end{document}